\documentclass[superscriptaddress,aps,preprintnumbers,amsmath,showpacs,amssymb,prd,nofootinbib,preprint]{revtex4-1}
\pdfoutput=1
\usepackage{booktabs} 
\usepackage{array} 
\usepackage[table,xcdraw]{xcolor} 
\usepackage{amsmath,amssymb}
\usepackage{graphicx}
\usepackage{fancyhdr}
\usepackage{bm, color}
\usepackage{slashed,amsthm,amsfonts,empheq}
\usepackage[caption=false]{subfig}
\usepackage{hyperref}
\usepackage{booktabs}
\usepackage{multirow}

\newcommand{\Slash}[1]{{\ooalign{\hfil#1\hfil\crcr\raise.167ex\hbox{/}}}}

\newcommand{\beq}{\begin{equation}}  \newcommand{\eeq}{\end{equation}}
\newcommand{\bef}{\begin{figure}}  \newcommand{\eef}{\end{figure}}
\newcommand{\bec}{\begin{center}}  \newcommand{\eec}{\end{center}}
  
\newcommand{\laq}[1]{\label{eq:#1}}  

\newcommand{\Eq}[1]{Eq.(\ref{eq:#1})}

\newcommand{\eq}[1]{(\ref{eq:#1})}

\newcommand{\SU}[1]{{\rm SU{#1} } }

\def\({\left(}
\def\){\right)}
\def\dt{\frac{d}{dt}}

\def\O{\mathcal{O}}

\newcommand{\OR}{~{\rm or}~}
\newcommand{\AND}{~{\rm and}~}

\newcommand{\GEV}{{\rm GeV}}

\def\e{\epsilon}
\def\f{\phi}

\def\k{\kappa}

\def\m{\mu}
\def\n{\nu}

\def\s{\sigma}

\def\G{\Gamma}

\def\L{\Lambda}

\def\tl{\tilde}
\def\*{\dagger}

\begin{document}
%
%
%
%
\preprint{TU-1287}

\title{
Axion Cosmology with Multi-Branch Yang-Mills
}


\author{Tsubasa Sugeno}
\affiliation{Department of Physics, Tohoku University, Sendai, Miyagi 980-8578, Japan}

 \author{Wen Yin}
\affiliation{Department of Physics, Tokyo Metropolitan University, Minami-Osawa, Hachioji-shi, Tokyo 192-0397, Japan}

\begin{abstract}
It is known that Yang-Mills theories, especially in the large-$N$ limit, exhibit a
$\theta$-vacuum structure with a multi-branched vacuum energy.
In this work, we demonstrate that this multi-branch structure can play a crucial role in
axion cosmology when the axion acquires its mass from the Yang-Mills sector, even when
that sector is never reheated by the inflaton.
The key observation is that the axion potential is directly tied to the tunneling rate
between adjacent branches. We find qualitatively new phenomena, including a new class of first-order phase transitions,
bouncing bubbles, and nested ``bubbles-within-bubbles.’’
When the axion has a decay constant around the Planck scale, as motivated by the string Axiverse, the axion can be driven near the hilltop by inflationary dynamics, allowing the phase transition to be triggered.  
The associated energy release can be large enough to generate a significant stochastic
gravitational-wave background, produce primordial black holes, or populate the Yang-Mills
sector with particles. These phenomena represent novel predictions of the Axiverse and should be taken into
account when assessing the cosmological impact of axions or axion-like particles.
To recover conventional axion cosmology, one must suppress or avoid the dynamics discussed in this work.
\noindent
\end{abstract}

\maketitle


\section{Introduction}

Axions and axion-like particles (ALPs) naturally appear as pseudo-Nambu-Goldstone bosons associated with spontaneously broken shift symmetries~\cite{Jaeckel:2010ni,Ringwald:2012hr,Arias:2012az,Graham:2015ouw,Marsh:2015xka,Irastorza:2018dyq,DiLuzio:2020wdo}. 
A well-known example is the QCD axion, whose coupling to the QCD topological term resolves the strong CP problem through the Peccei-Quinn (PQ) mechanism~\cite{Peccei:1977hh,Peccei:1977ur,Weinberg:1977ma,Wilczek:1977pj}. 
More generally, axions are expected to couple to various Yang-Mills sectors in ultraviolet (UV) completions such as string theory, where numerous axion fields with different decay constants and couplings can coexist-the so-called \emph{Axiverse}~\cite{Arvanitaki:2009fg,Acharya:2010zx}.
The effective theory of an axion field~$\phi$ can be characterized as a CP-odd scalar (assuming CP is a good symmetry) possessing an exact discrete shift symmetry,
\beq
\phi \;\to\; \phi + 2\pi f_{0},
\eeq
where $f_{0}$ denotes the axion decay constant.

In such multi-axion frameworks, each axion may couple to a distinct non-Abelian gauge group, giving rise to a $\theta$-vacuum structure with multiple branches in the effective potential. From the dimensional transmutation, the axion mass scale distributes over an exponentially wide range. 
While this multi-branch structure is an inherent property of confining Yang-Mills dynamics, especially in the large-$N$ limit~\cite{Witten:1980sp,Witten:1998uka}, it is often neglected in phenomenological models, where the potential is approximated by a single cosine function. Nevertheless, the existence of multiple branches can qualitatively change the nonperturbative vacuum dynamics, particularly in the early universe.

A plausible feature of the Axiverse is that the inflaton for cosmic inflation may not reheat all sectors uniformly. 
In such a case, many hidden Yang-Mills sectors remain cold and unpopulated after reheating, which helps explain why the Universe is not overclosed by dark-sector matter or radiation.

In this work, we point out the importance of the multi-branch structure of the Yang-Mills potential in axion cosmology, even when the Yang-Mills sector is not reheated by the inflaton. The key observation is the direct relation between the axion potential and the tunneling rate between adjacent branches.
If the axion is set near the hilltop, which can be naturally realized by inflationary selection for string axions, its post-inflationary evolution can trigger a strongly first-order phase transition, characterized by ultra-relativistic bubbles, inside which further bubble nucleation can occur-forming “bubbles within bubbles.’’\footnote{In another scenario irrelevant to the multi-branch structure of the axion potential, the production of enormous axion particles with sufficiently low momenta can generically induce the formation of domain wall loops inside domain wall loops, as shown by lattice simulations in Refs.\,\cite{Narita:2025jeg,Miyazaki:2025tvq}.}
On the other hand, even when the axion is not driven to the hilltop, bubble formation can still occur. In this case, the nucleation events may be rare, and as the potential bias disappears once the axion begins oscillating around the degenerate vacua of different branches, the resulting bubbles can bounce due to the wall tension. 

If the Yang-Mills vacuum energies are not too small compared with the critical energy density at the transition, 
the resulting phenomena can generate a significant stochastic gravitational-wave background, produce large overdensities in the Universe, lead to the formation of primordial black holes, and induce particle production within the Yang-Mills sector.

When the axion has a fundamental decay constant of $f_0=10^{16\text{-}17}\,\GEV$, as motivated by the string Axiverse, the energy density associated with the transition is expected to be large thanks to the enhanced effective decay constant $f_\phi=Nf_0$ in a single branch.
Thus, these effects provide a novel probe of hidden Yang-Mills sectors in the Axiverse and highlight the importance of properly accounting for the multi-branch structure in realistic axion cosmology. 
We also comment that similar dynamics can occur even without introducing an axion.

Let us mention some related topics. The $\theta=\pi$ domain wall in Yang-Mills theory, associated with spontaneous CP violation, has been studied since the early works on the Dashen phenomenon~\cite{Dashen:1970et,Baluni:1978rf} and in large-$N$ analyses of QCD~\cite{Witten:1980sp,DiVecchia:1980yfw}. More recently, the structure of the $\pi$ domain wall has been revisited in modern approaches~\cite{Gaiotto:2017yup,Komargodski:2017dmc}. Domain walls are also expected to form in the post-inflationary PQ breaking scenario, where the axion field satisfies $\theta \sim \pi$ around cosmic strings~\cite{Gabadadze:2000vw,Sikivie:1982qv}. Induced domain walls generated by other domain walls or bubble walls, as well as axion waves, have been studied in Refs.~\cite{Lee:2024xjb,Lee:2024oaz} (see also \cite{Lee:2024toz, Lee:2025zpn} for the relevant study with the induced wall in the context of stability of the string bundle, and transient potential bias due to axion motion). Such induced domain walls can effectively be interpreted as an axion-mediated force~\cite{Kim:2021eye}. In these studies, the relevant particles-those involved in domain-wall or bubble-wall dynamics-are typically assumed to be thermally populated or to appear in post-inflationary scenarios. In this work, however, we consider a pre-inflationary scenario in which the Yang-Mills sector is not reheated and various phenomena appear due to the multi-branch structure. Thus the phenomena are naturally realized in a minimal axion/ALP model.

Ultra-relativistic bubble walls from strongly first-order phase transitions are of considerable interest~\cite{Coleman:1977py,Coleman:1980aw,Guth:1982pn}, 
both from the viewpoint of their intrinsic dynamics and because of their implications for enhanced gravitational-wave signals~\cite{Bodeker:2009qy,Bodeker:2017cim,Hoche:2020ysm,Azatov:2020ufh,GarciaGarcia:2022yqb,Azatov:2024crd,Espinosa:2010hh,Caprini:2019egz}. 
In conventional thermal phase transitions, realizing a strongly first-order transition with bubble nucleation typically requires a special setup~\cite{Konstandin:2011dr,Iso:2017uuu,Azatov:2023xem}, 
since one must obtain sufficiently large latent energy while simultaneously suppressing the friction on the bubble walls, both of which are controlled by similar microscopic couplings.
In our (hilltop-axion) scenario, by contrast, ultra-relativistic bubbles arise naturally: the friction force is absent, while the axion dynamics can trigger tunneling with a large latent energy.

\section{Yang-Mills vacuum structure and Tunneling rate}

We consider the $\SU(N)$ Yang-Mills theory.
Before introducing the axion coupling, let us recall that a Yang-Mills sector without any axionic degree of freedom can exhibit a nontrivial vacuum structure characterized by the topological $\theta$ term,
\beq
{\cal L}_\theta \;=\; \frac{\theta}{32\pi^2}\, G_{\mu\nu}^a\,\tilde{G}^{a\,\mu\nu},
\eeq
where $G_{\mu\nu}^a$ is the field strength and $\tilde{G}^{a\,\mu\nu}$ its dual. 
At zero temperature, the $\theta$ parameter labels distinct topological sectors of the gauge theory, and the physical vacuum energy (density) $E(\theta)$ is a periodic function of~$\theta$ with period $2\pi$.

\begin{figure*}[!t]
    \begin{center}
     \includegraphics[width=0.7\textwidth]{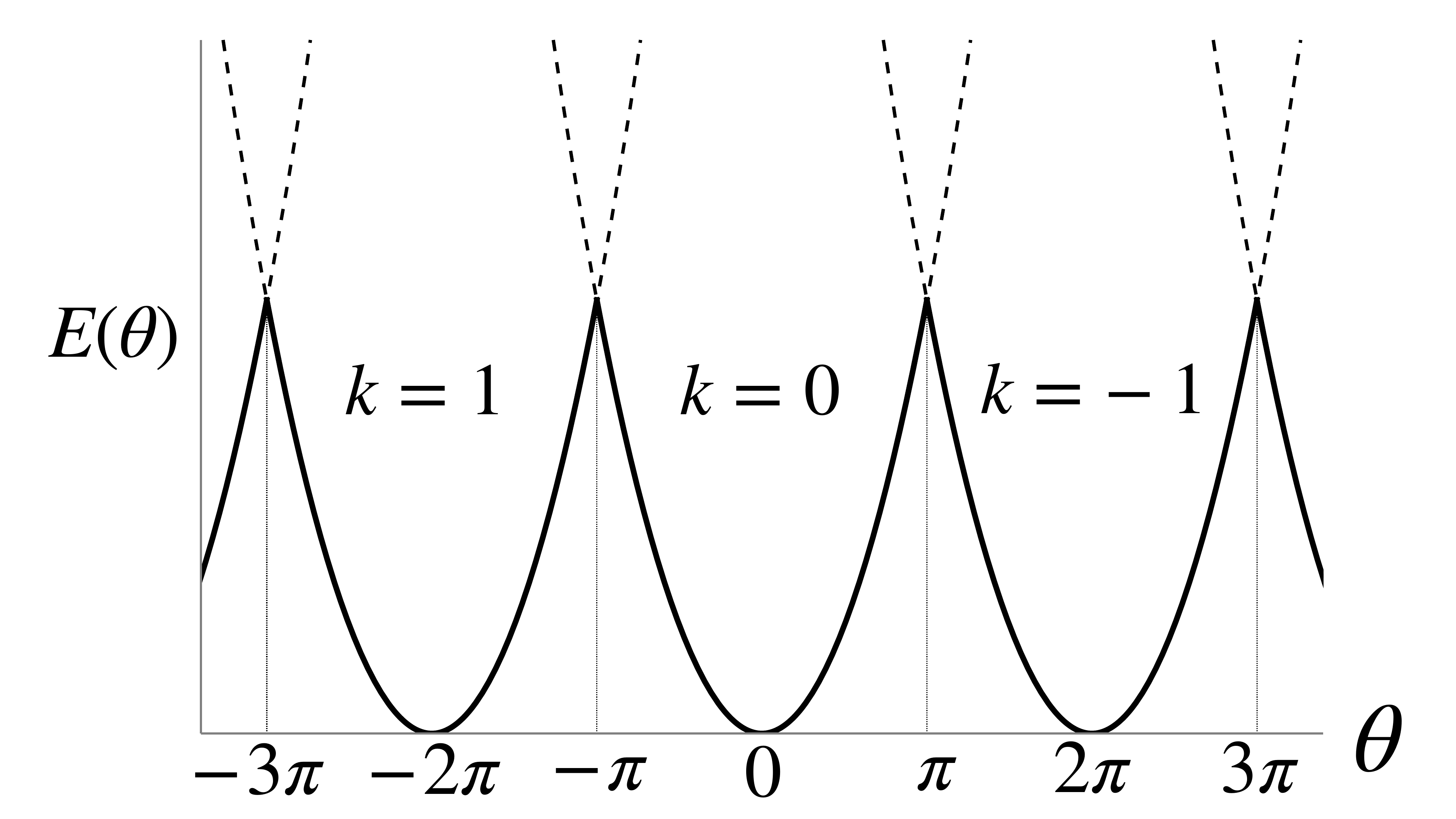}
    \end{center}
    \caption{Schematic picture of the vacuum energy $E(\theta)$. 
Each smooth curve (composed of solid and dashed segments) corresponds to a vacuum labeled by an integer $k$. 
The solid curve represents the true vacuum, while the dashed curves represent metastable vacua. 
The $2\pi$ periodicity of $\theta$ is realized through the presence of multiple branches.
    }
    \label{fig:multibranch}
\end{figure*}

Large-$N$ arguments~\cite{Witten:1980sp,Witten:1998uka} suggest that the true vacuum energy is a multi-branched function of $\theta$, schematically given by
\beq
E(\theta) \;=\; N^2 \Lambda^4\,\min_k\, h\!\left(\frac{\theta + 2\pi k}{N}\right),\laq{etheta}
\eeq
where $\Lambda$ denotes the characteristic confinement scale of the gauge dynamics, and $h(x)$ is some function.
The important point is that $E(\theta)$ is not a smooth $2\pi$-periodic function of $\theta$; rather, the $2\pi$ periodicity is realized only after combining different vacua labeled by an integer $k$.
Fig.~\ref{fig:multibranch} shows a schematic picture of $E(\theta)$.
Although the above discussion is based on the large-$N$ limit, it is believed that this multi-branch structure also appears at finite $N$. 
For example, recent numerical simulations designed to tackle the sign problem provide evidence of spontaneous CP breaking at $\theta=\pi$ even for finite $N$, such as $N=2$~\cite{Kitano:2020mfk,Kitano:2021jho,Yamada:2024vsk,Yamada:2024pjy,Hirasawa:2024fjt}.
More abstractly, new ’t~Hooft anomalies have been identified that forbid a smooth $2\pi$-periodic vacuum~\cite{Cordova:2019jnf,Cordova:2019uob} (see also~\cite{Gaiotto:2017yup,Kikuchi:2017pcp}).

This branch structure can be more easily understood with a single flavor quark {following \cite{Witten:1980sp}}. 
\beq
{\cal L}= -\frac{1}{4g^2} G^a_{\m\n}G^{a\m\n} + \bar{\psi}(i\Slash{D}-m_q)\psi
\eeq 
where we removed the $\theta$ term and we use the $m_q$ as conplex parameter $m_q=|m_q|e^{i\theta}.$
The anomalous $U(1)$ phase corresponds to the massive $\eta'$ messon.

The chiral effective Lagrangian is built from a unitary field
$U(x) = e^{i\eta'(x)/F_{\eta'}} \in U(1)$.
Owing to the anomaly, in addition to
\beq
{\cal L}_{q} = \frac{1}{2}\kappa_1 (m_q U + \text{h.c.})
  = \kappa_1 |m_q| \cos\!\left(\frac{\eta'}{F_{\eta'}} + \theta\right),
\eeq
there are terms that remain finite in the limit $|m_q| \to 0$,
\beq
\delta{\cal L} = F[\eta'] = \frac{\chi}{2} \left(\frac{\eta'}{F_{\eta'}}\right)^2 + \cdots .
\eeq
Here $\kappa_1$ and $\chi$ are dimension-3 and dimension-4 parameters of order the dynamical scale, 
and $F$ denotes a function generated by nonperturbative effects.  
When the mass term from $F[\eta'] \simeq \chi\,\eta'^2/(2F_{\eta'}^2)$ is suppressed-e.g., in the large-$N$ limit-one obtains multiple vacua of $\eta'$ from ${\cal L}_q$, with a small modulation coming from $\delta{\cal L}$.  {The vacuum of $\eta'$ potential can be determined by $\eta'_{k,\rm min}\sim -(2 \pi k +\theta)F_{\eta'}$. }
Integrating out $\eta'$ around each vacuum leads to the effective vacuum energy,
\beq
E[\theta] \simeq \min_{k} F\!\left[-(\theta + 2\pi k)\,F_\eta'\right],
\eeq
thereby reproducing the form in Eq.~\eq{etheta}.

Deriving the shape of $F$ is beyond the scope of this work, and we assume the shape of $h$ may have the hilltop and bottom close to both of which are controlled by quadratic terms. 

The excited false vacuum undergoes tunneling-induced decay into an adjacent vacuum\footnote{When the decay rate is highly suppressed the leading contribution becomes the decay into non-adjacent vacua. This effect is not taken into account in this paper.}. 
The tunneling rate can be estimated with 
\beq 
\G \sim \L^4 e^{-S_4[\theta]}.
\eeq 
With the large $N$ limit, one can use the thin wall approximation to get  $S_4\propto \frac{\s^4}{\e^3}$ where $\e$ is the energy difference controlled by \beq \epsilon=|F[\eta'_{k,\rm min}]-F[\eta'_{k\pm 1,\rm min}]|\simeq 2\pi F_{\eta'}|F'[\eta'_{k,\rm min}]|.\eeq
$\sigma \sim (\k_1 m_q)^{1/2} F_{\eta'}$ is the tension of the bubble wall.  
$S_4 \propto 1/(|F'[\eta'_{k,\rm min}]|)^3\propto 1/|h'|^3$ thus it is inversely proportional to the derivative of the vacuum energy in a single branch with respect to $\theta$. In particular, close to the hilltop or minimum of $F$ (or $h$), the tunneling rate gets exponentially suppressed. As we will consider $F$ (or $h$) as the axion potential in the next section, the axion dynamics will be linked to the tunneling rate. \\

So far, we have used the simple chiral perturbation theory as a concrete example. 
Similar phenomena also arise in softly broken $\mathcal{N}=1$ super Yang-Mills theory without relying on the large-$N$ argument~\cite{Shifman:1987ia,Dvali:1996xe,Shifman:1998if,Yonekura:2014oja}.  
For $\mathcal{N}=1$ super Yang-Mills with a gluino mass and gauge group $G=\SU(N)$, one explicitly finds $N$ distinct vacua, yielding a vacuum energy {in a single branch} 
$\propto 1 - \cos\!\left[(\theta + 2\pi k)/N\right]$ for $k=0,1,\ldots,N-1$, which again form the same type of multi-branch structure discussed above.

In non-supersymmetric Yang-Mills theory, deriving general analytic results is more challenging.  
{Standard Model QCD is considerred to be unlikely to exhibit a multi-branch structure, 
as light chiral quarks tend to smooth out the would-be Yang-Mills branches.}
Nevertheless, as mentioned above, numerical simulations and various theoretical arguments strongly support the existence of this multi-branch vacuum structure {of generic Yang-Mills theories.}

\begin{figure*}[!t]
    \begin{center}
     \includegraphics[width=0.9\textwidth]{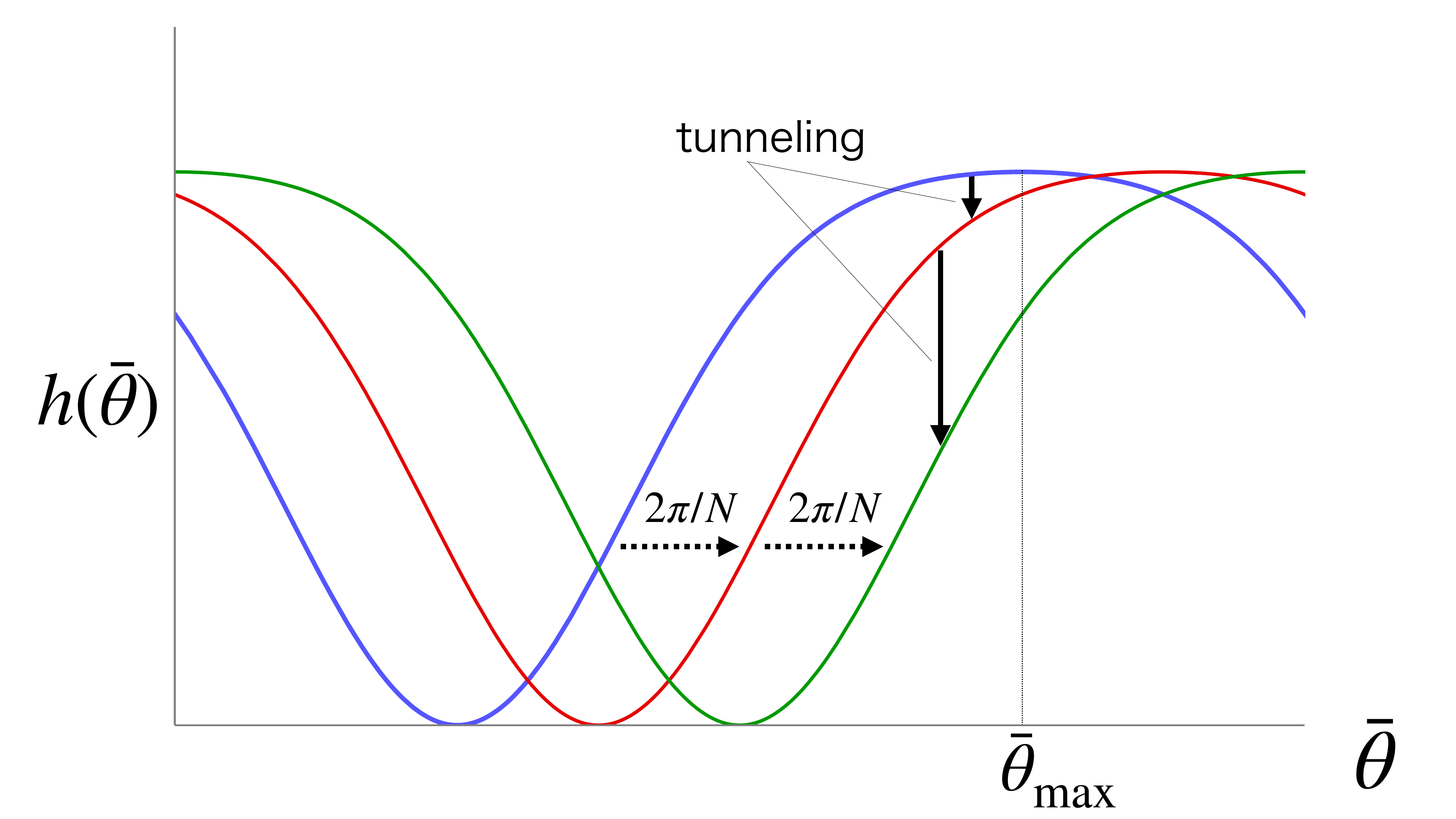}
    \end{center}
    \caption{Schematic picture of the tunneling near the hilltop.
    Blue, red, and green lines show the different vacuum branches.
    The tunneling occur between neighboring branches.}
    \label{fig:tunneling_hiltop}
\end{figure*}

{The multi-branch behavior and the resulting scaling $S_4 \propto (\Delta h)^{-3}$ may be quite generic, 
with $\Delta h$ denoting the vacuum-energy difference between adjacent branches.}
At least when the wall tension has no significant $\theta$-dependence, the form of Eq.~\eq{etheta} 
implies the same scaling. This behavior is illustrated in Fig.~\ref{fig:tunneling_hiltop}, 
where a generic function $h$ is assumed. Near the hilltop or the minimum, the difference in $h$ becomes small, 
which in turn suppresses the tunneling rate.

In the following unless otherwise stated, without loss of generality, we focus on the $k=0$ branch is excited and consider the decay 
into the adjacent branch ($k=-1$) within the range $\bar{\theta}>0$:
\beq
S_4 = \frac{(2\pi/N)^3\,\bar{S}}
           {|h(\bar{\theta}) - h(\bar{\theta}-2\pi/N)|^3}
      \;\simeq\;
      \frac{\bar{S}}{|h'(\bar{\theta})|^3},
\qquad
\bar{\theta}\equiv \frac{\theta}{N}. \laq{S4}
\eeq
The dimensionless constant $\bar{S}$ encodes the details of the bubble-wall tension, 
the overall scale of the potential, and numerical factors. 
Importantly, through the function $h$, the tunneling rate is directly linked to the $\theta$ (and thus axion) potential, 
which is a key observation of this work.  
For simplicity, in the last approximation, we adopt the large-$N$ limit, which will also be used in the numerical estimates in later sections.

{We also comment that a very large $N$ suppresses the tunneling rate. 
For instance, in the large-$N$ limit of softly broken pure $\mathcal{N}=1$ super Yang-Mills theory with a not-too-small gluino mass $m_\lambda$, the bounce action behaves as 
$\bar{S} \sim 10^{-6}\frac{N\Lambda^3}{(m_\lambda/(g^2 N))^3}$ for $\bar{\theta}=\mathcal{O}(1)$~\cite{Shifman:1998if,Yonekura:2014oja}.}
Although we focus on the large-$N$ limit in the following for analytical and illustrative purposes, 
we emphasize that the phenomena discussed in this work persist qualitatively as long as the theory exhibits a multi-branch potential and 
$h[\bar\theta]-h[\bar\theta-2\pi/N]$ shows a similar behavior. 
For instance, even for moderately small $N$, tunneling near the hilltop is suppressed because the mass difference between adjacent branches becomes small (see the crossing between the adjacent branches in Fig.~\ref{fig:tunneling_hiltop}.).

\section{Inflationary selection of the excited vacuum and the late-time cosmology without an axion}

Let us discuss the cosmology of the Yang-Mills theory with the multi-branch structure. 
In this section, we do not introduce axion and consider a fixed $\theta$. 
Suppose the cosmic inflation occurs with an expansion rate $H_{\rm inf}$. If $\min_k \G[(\theta+2\pi k)/N ] \gg H_{\rm inf}^4$, during the inflation all the branches are settles into the true vacuum. 
Then our Universe is in the branch with the minimum energy with the potential \Eq{etheta}. Alternatively if $H_{\rm inf}^4\gg \L^4$, the Gibbons-Hawking temperature is higher than the Yang-Mills confining scale, and the Yang-Mills sector is populated during inflation. The time-evolution is similar to the thermalized Yang-Mills sector. 

Thus, throughout the paper, we assume 
\beq
\max_k \G[(\theta+2\pi k)/N] \ll H_{\rm inf}^4, \AND ~~H^4_{\rm inf}  \ll \L^4  \laq{cond}
\eeq 
during the cosmic inflation. Again we assume, $k=0$ has the  largest decay rate without the loss of generality. 
We will show that even in the limit that the Yang-Mills sector is never reheated, the false vacuum for a given $\theta$ can be populated due to cosmic inflation and undergoes the tunneling in the post inflationary Universe.

To illustrate this, let us consider a two-state system consisting of an excited and a ground state, where (pure de Sitter) inflation begins with volumes $V_{i,e}$ and $V_{i,g}$\footnote{One may consider, $V_{i,e}\sim V_{i,g}$ if they are in thermal equilibrium with a high enough temperature before the inflation.}, respectively. 
The energy difference between the two states is assumed to be $\Delta V = \tilde{c}\,\Lambda^4$, 
with $\tilde{c}$ being an $\mathcal{O}(1)$ model-dependent constant. For the adjacient branch and for the large $N$ limit, $\tilde{c}\sim N |h'| 2\pi.$
The physical volume remaining in the excited state at cosmic time $t$ evolves as
\beq
V_{e}(t)= V_{i,e}\, \exp\!\left[3t\!\left(H_{\rm inf,e} - c\,\Gamma\,H_{\rm inf,e}^{-3}\right)\right],
\eeq
where $c$ is an $\mathcal{O}(1)$ parameter and $H_{\rm inf,e}$ denotes the Hubble parameter of the excited state. 
The second term in the exponent represents the effective tunneling rate per Hubble patch. 

The ground-state volume then evolves as
\begin{align}
V_{g}(t)
&= V_{i,g}\, e^{3t H_{\rm inf,g}}
+ \int_{t_d}^{t}\! dt'\, e^{3H_{\rm inf,g}(t-t')}\,\frac{3 c\,\Gamma}{H_{\rm inf,e}^3}\,V_{e}(t') \nonumber\\[4pt]
&= e^{3t H_{\rm inf,g}}\!\left(
V_{i,g}
+V_{i,e}\,
\frac{c\,\Gamma\,H_{\rm inf,e}^{-3}}
{c\,\Gamma\,H_{\rm inf,e}^{-3}- (H_{\rm inf,e}-H_{\rm inf,g})}
\right)
- V_{e}(t)\!\left(
\frac{c\,\Gamma\,H_{\rm inf,e}^{-3}}
{c\,\Gamma\,H_{\rm inf,e}^{-3}- (H_{\rm inf,e}-H_{\rm inf,g})}
\right).
\end{align}
In the long-time limit we obtain
\beq
\lim_{t\to \infty}\frac{V_e(t)}{V_g(t)} \gg 1,
\qquad {\rm if} \qquad
\left(H_{\rm inf,e} - \frac{c\,\Gamma}{3H_{\rm inf,e}^{3}}\right) > H_{\rm inf,g}.
\eeq
This means that, for sufficiently long inflation, the Universe can remain in the excited branch provided that
\beq
\label{sbarcond}
\frac{\tilde{c}\,\Lambda^4}{6H_{\rm inf,e}M_{\rm pl}^2} > 
\frac{c\,\Gamma}{3H^3_{\rm inf,e}}
\;\;\Longrightarrow\;\;
\frac{\tilde{c}}{c}\,
\frac{H_{\rm inf,e}^2}{2M_{\rm pl}^2}
> 
e^{-S_4}.
\eeq
Here $M_{\rm pl}\approx 2.4\times 10^{18}\,\GEV$ is the reduced Planck scale, which is much larger than $H_{\rm inf,e,g}$, and we have used the expansion
$H_{\rm inf,g}\simeq H_{\rm inf,e}- \tilde{c}\,\Lambda^4/(6 H_{\rm inf,e} M_{\rm pl}^2)$.
For instance, taking $H_{\rm inf,e}=10^{13}\,{\rm GeV}$, this condition is satisfied for $\bar{S}/h'^3\simeq 25$. 

The inflationary duration required for the excited state to become exponentially favored can be estimated as
\beq 
\Delta t \gg  \frac{1}{-c\,\Gamma H_{\rm inf,e}^{-3}+\,\tilde{c}\,\Lambda^4/(6H_{\rm inf,e}M_{\rm pl}^2)}
\simeq \frac{6H_{\rm inf,e}M_{\rm pl}^2}{\,\tilde{c}\,\Lambda^4},
\eeq 
where, in the last expression, we have neglected the decay rate. 
For $H_{\rm inf,e}=10^{13}\,{\rm GeV}$ and $\Lambda\sim 10^{15}\,\GEV$, the corresponding number of $e$-folds is of order $10^{3}$.\footnote{With a non-negligible constant term compared to the inflaton potential, one can also rescue natural inflation~\cite{Murase:2025uwv}.} 
Thus, we do not need to invoke eternal inflation as long as $\Lambda$ is not extremely small, and we do not address the measure problem.

Under the assumption of a fixed $\theta$ in this section, satisfying Eq.~(\ref{sbarcond}) implies that the inflationary dynamics can select the excited state. 
After inflation ends, when the tunneling rate becomes comparable to the fourth power of the Hubble expansion rate, $\Gamma= H^4(t_n)$, 
bubbles of the lower-energy branch nucleate and expand, leading to vacuum tunneling.
We define this nucleation time $t=t_n$. 
The corresponding bubble walls are composed of non-Abelian field configurations interpolating between two neighboring $\theta$ branches (in the discussed chiral perturbation theory it is the $\eta'$ field).

If the excited-state energy density does not dominate the Universe when $\Gamma\sim H^4$ and $H$ begins to redshift, 
a ``first-order phase transition’’ can indeed take place. 
If, instead, the false-vacuum energy dominates, the transition may not complete, leading to false-vacuum (old) inflation.  
Such a completed transition could in principle release latent energy $\Delta E \sim \tl c\Lambda^4  $. 
Since we consider that the Yang-Mills sector rarely couples to the existing particles in the Universe, the friction force is highly suppressed. Thus this bubble expands until it collides with other bubbles. Since the tunneling rate is a constant value, the typical bubble size at the transition is of order $1/H$. 
If it occurred during a cosmological epoch, it would be capable of sourcing a stochastic gravitational-wave background due to the bubble collistion.

To sum up, Table~\ref{tab:cond} summarizes the possible inflationary and post-inflationary histories of the two state system, where for simplicity we take $H_{\rm inf}\simeq H_{\rm inf,e}$.  
We also use the fact that the Hubble parameter after inflation is smallest today, and therefore the condition $\Gamma > H_0^4$ ensures that tunneling can occur by the present epoch.  
We see that in case (I) late-time tunneling is possible, allowing us to probe the dark-sector scenario.  
In case (II), although the system remains in the false vacuum, the cosmological constant can be fine-tuned to zero, and the early cosmology becomes indistinguishable from the standard one.

\begin{table}[t]
\centering
\caption{Inflationary excitation and (post-inflation) tunneling conditions.}
\label{tab:cond}
\vspace{4pt}
\begin{tabular}{c| c| c| c}
\hline
\textbf{Case} & \textbf{Condition 1} & \textbf{Condition 2} & \textbf{Tunneling until now}\\[3pt]
\hline
(I) & 
$\displaystyle \frac{\tilde{c}\Lambda^4}{6H_{\rm inf}M_{\rm pl}^2} \;>\; \frac{c\,\Gamma(\theta)}{3H_{\rm inf}^3}$  
& 
$\displaystyle\ \frac{\Gamma(\theta)}{H_0^4}\ \ge 1$
& Yes \\[10pt]
(II)   & 
$\displaystyle \frac{\tilde{c}\Lambda^4}{6H_{\rm inf}M_{\rm pl}^2} \;>\; \frac{c\,\Gamma(\theta)}{3H_{\rm inf}^3}$ 
& 
$\displaystyle \ \frac{\Gamma(\theta)}{H_0^4}\ <\ 1$
& No \\[10pt]
(III) & 
$\displaystyle \frac{\tilde{c}\Lambda^4}{6H_{\rm inf}M_{\rm pl}^2} \;< \; \frac{c\,\Gamma(\theta)}{3H_{\rm inf}^3}$ &
-- & No \\[3pt]
\hline
\end{tabular}
\end{table}
\medskip
\noindent

\textbf{Bubbles within bubbles with multiple branches}

So far, we have focused on the two-state system for simplicity.  
In more general cases, multiple tunneling channels exist.  
If inflation populates the branch with the highest energy, the system may subsequently undergo a cascade of tunnelings in late-time cosmology.

{Interestingly, for sufficiently large $N$, there always exists a branch arbitrarily close to the hilltop, where the decay rate is highly suppressed, since the nearest distance to the hilltop is bounded by $2\pi/N$. The most excited branch is naturally selected.}

When $\bar{\theta}$ lies near the hilltop value $\bar{\theta}_{\rm max}$, we have
\beq
h \simeq \frac{h''(\bar\theta_{\rm max})}{2}\,(\bar{\theta}-\bar{\theta}_{\rm max})^2 .
\eeq
If this branch is populated during inflation, the associated bounce action scales as
\beq 
S_4 \propto |h'|^{-3} \propto \frac{1}{|\bar\theta - \bar\theta_{\rm max}|^3}.
\eeq
If the post-inflationary tunneling condition is satisfied, the vacuum value shifts to
\beq 
h(\bar{\theta}\pm 2\pi/N)\simeq 
\frac{h''(\bar\theta_{\rm max})}{2}\,(\bar\theta\pm 2\pi/N - \bar\theta_{\rm max})^2 ,
\eeq
under the assumption that $2\pi/N$ is small enough for the expansion to be valid but larger than $|\bar{\theta}-\bar\theta_{\rm max}|$.\footnote{Even for not-too-small $2\pi/N$, we expect the argument to hold if $\bar\theta\sim \bar\theta_{\rm max}$ is sufficiently suppressed compared to the other branch (see Fig.\ref{fig:tunneling_hiltop}).}  
In this case $|h'|$ increases, further suppressing $S_4$ and making subsequent tunneling even more efficient.

In this scenario, once a primary bubble forms, the $\theta$ in the branch inside the bubble no longer stays near the hilltop but moves to a region where $|h'|$ is larger. 
Since the bounce action scales as $S_4 \propto |h'|^{-3}$, the tunneling rate becomes 
significantly higher in the interior than in the exterior, causing additional transitions 
to occur rapidly.  
As a result, many secondary bubbles can nucleate inside the primary bubble.

To see this more clearly, let us perform a simple estimate by focusing on the 
secondary-bubble formation inside a primary bubble of radius $R$.  
We neglect bubble-bubble collisions and assume that all bubbles expand at the speed of light.  
For illustration, we consider $d$ spatial dimensions, so that the local tunneling probability per unit time 
in a small interior region scales as $r^{d-1}dr\,d\Omega_d$ in polar coordinates measured 
from the center of the primary bubble, where $\Omega_d$ denotes the $d$-dimensional solid angle.

\begin{figure*}[!t]
    \begin{center}
     \includegraphics[width=0.8\textwidth]{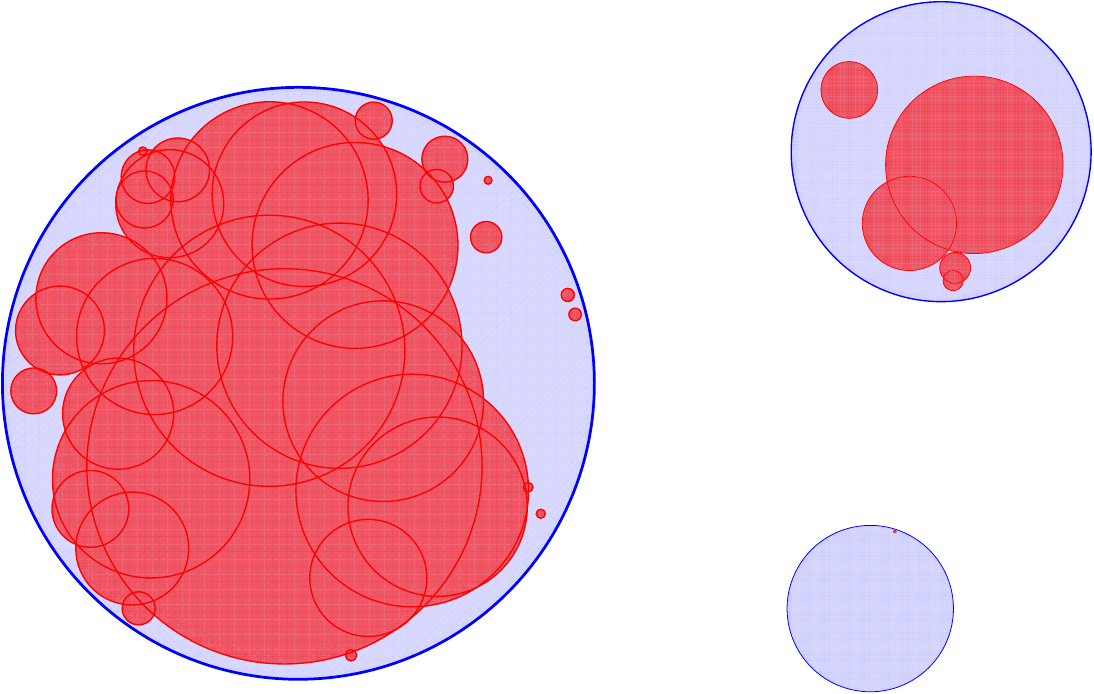}
        \vspace{-5mm}
    \end{center}
    \caption{A simple 2D simulation of the ``bubbles-within-bubbles'' scenario, assuming an additional cascade of tunneling events. The blue (red) circles denote the primary (secondary) bubbles.}
    \label{fig:binb}
\end{figure*}

For a primary bubble of radius $R$, a secondary bubble nucleated at radius $r$ can only 
appear during the time interval of length $R-r$ before the primary wall reaches $R$.  
Taking into account the subsequent expansion of the secondary bubble, the radii of the 
secondary bubbles are uniformly distributed in the interval $[0,\,R-r]$.  
Using these ingredients, we stochastically generate the distribution of bubbles inside a 
primary bubble of radius $R$.

The result of this simple simulation with $d=2$ is shown schematically in 
Fig.~\ref{fig:binb}, for three different values of $R$ and a fixed nucleation rate for 
secondary bubbles.  
In the figure, the final vacuum surrounded by the secondary bubbles is indicated in red, 
while the intermediate vacuum enclosed by the primary bubble is shown in blue.

One finds that for sufficiently large primary bubbles, many secondary bubbles are produced.  
Secondary bubbles tend to be larger when they nucleate earlier and closer to the center of 
the primary bubble, whereas their production is suppressed for smaller primary bubbles due 
to the limited interior volume and shorter available time.  
The expected number of secondary bubbles inside a primary bubble of radius $R$ scales as 
$R^{d+1}$.

So far we have focused on the three-state system.  
With more branches, however, more cascade transitions can occur.  
The branch inside a secondary bubble moves even farther away from the hilltop, 
so that $|h'|$ increases further and the bounce action decreases.  
This again enhances the tunneling rate and allows additional bubble nucleation inside the 
secondary bubbles.  
The process continues until the branch inside the innermost bubble enters a region where 
$|h'|$ becomes sufficiently small and tunneling effectively shuts off.

Collisions, which we have neglected in the simulation, among the inner bubbles can produce Yang-Mills particles, which eventually 
act as a source of friction for bubbles nucleating at later stages.

The resulting gravitational-wave spectrum from such multi-stage collisions may exhibit 
distinctive features, making this scenario an interesting target for further investigation.

\section{Impacts of Multi-Branch Yang-Mills dynamics on Axion
Cosmology}

We now consider the case where
\beq
\theta = \frac{N\phi}{f_\phi}
\eeq
is identified with the axion field. 
The factor of $N$ is our convention.
We note that \beq 
f_\f = N f_{0}.
\eeq
Thus the effective decay constant $f_\f$ is larger than the one from the periodicity $f_{0}$. 
This enhancement was considered to induce the superplanckian effective decay constant and field excursion~\cite{Kaloper:2011jz,Dubovsky:2011tu,Yonekura:2014oja}. Notice that if the axion is motivated from string theory $f_0\sim 10^{16-17}\GEV$, it is not difficult to have  $f_\f\sim M_{\rm pl}$ with $N\gtrsim \O(1-10)$. 

Within a single branch, the axion potential is generated nonperturbatively by the Yang-Mills dynamics as
\beq
V = N^2 \Lambda^4\, h(\phi/f_\phi).
\eeq
For instance, when $\phi$ couples to $\mathcal{N}=1$ super Yang-Mills theory with $G=\SU(N)$, $h(x)=1-\cos(x)$ has periodicity $\theta \to \theta +2\pi N$.

Suppose that the axion and the associated Yang-Mills sector are sequestered from the thermal bath of the Standard Model, i.e. very weakly coupled.
In this case, the PQ symmetry is never restored after inflation, and the Yang-Mills sector.

When $H_{\rm inf}^4\ll \min_{k}\Gamma({\bar\theta_i}+2\pi k/N)$, 
in the axion case, even if $\bar\theta_i=\O(1)$, the tunneling occurs during inflation and the axion starts to oscillate around the most close minimum in \Eq{etheta}.

Since we assume \Eq{cond},
the axion need not reside in the branch with the smallest energy at $\theta_i$ during inflation.

\subsection{Initial axion configuration from inflation}

After inflation driven by an inflaton field, the axion field is expected to take a nearly homogeneous value $\theta_i$ across our entire observable Universe.  
This value is determined by stochastic fluctuations of the axion field during and before inflation and corresponds to the standard ``pre-inflationary PQ breaking’’ scenario often invoked in the vacuum misalignment mechanism for explainning the dark matter~\cite{Preskill:1982cy,Abbott:1982af,Dine:1982ah}.

Focusing on the single branch whose minimum is $\theta=0$, 
for simplicity of our analysis we consider 
\beq 
\bar{\theta}_i \equiv \theta_i/N \lesssim 1 ~~\OR~~ \bar{\theta}_i \sim \bar\theta_{\rm max}. \laq{cases}
\eeq 
{Other than this region, we need to provide concrete $h$, while all the estimation can be performed similarly. }
Here we assume $\bar{\theta}_i \ge 0$ without loss of generality.\footnote{It should be noted that $N$ must be sufficiently large when considering tunneling with 
$\bar\theta_i \lesssim 1$ under the approximation \eq{vapp0}, 
since $-\pi < \theta_i < \pi$ corresponds to the true vacuum and no decay occurs in this range.  
With not too large $N$, given that $\theta_i>\pi$ and $\bar\theta_i \gtrsim 1$, the tunneling can still take place, but a detailed knowledge of the functional form of $h$ is required for a reliable discussion but $\bar \theta_i<1$.}

One can naturally have the relation depending on the inflationary or reheating dynamics. For instance, during prolonged inflation with $H_{\rm inf}^4 \lesssim N^2\Lambda^4$, the equilibrium distribution favors $\bar\theta_i \approx 0$~\cite{Graham:2018jyp,Guth:2018hsa,Ho:2019ayl,Alonso-Alvarez:2019ixv}, 
provided that the potential minimum does not shift between inflation and the post-inflationary epoch and the effective decay constant is subplanckian.  
If the minimum is displaced, for example through inflaton mixing, one can instead realize the hilltop condition, $\bar\theta_i \approx \pi$, for the case $h[x]=1-\cos[x]$ or intermediate values of $\mathcal{O}(1)$ depending on the UV completion~\cite{Daido:2017wwb,Co:2018mho,Takahashi:2019qmh,Takahashi:2019pqf,Nakagawa:2020eeg,Narita:2023naj}. \\

The hilltop axion can also be realized through inflationary selection, independently of the detailed form of the function $h$ in our setup, provided that the effective decay constant satisfies $f_\phi \sim M_{\rm pl}$. In contrast to the case discussed in the previous section, the same argument cannot be applied here because $\bar\theta$ can now slow-roll toward the minimum within a single branch.\footnote{One may alternatively introduce an additional axion mass with the usual periodicity during inflation, which disappears after reheating due to thermal effects. In that case, the previous discussion applies and the hilltop can again be selected with $f_\f \ll M_{\rm pl}$.}

For the moment for illustrative purpose, we assume that
\beq
\Gamma[\bar\theta] \ll H_{\rm inf}^4
\eeq
for all values of $\bar\theta$, so that tunneling effects can be neglected. We will later relax this assumption and include their impact.

To study this, let $L[\phi,t]$ denote the comoving-volume distribution satisfying the 
Fokker-Planck equation~\cite{Starobinsky:1986fx,Starobinsky:1994bd,Nakao:1988yi,Nambu:1988je,Nambu:1989uf,Linde:1993xx}:
\beq\laq{FP}
\partial_t L[\phi,t]
= 3\,(H[\phi]-H_{\rm inf})\,L[\phi,t]
  + \partial_\phi\!\left[
      \frac{V_{,\phi}}{3H[\phi]}\,L[\phi,t]
      + \frac{H[\phi]^{3/2}}{8\pi^2}\,
        \partial_\phi\!\bigl(H[\phi]^{3/2} L[\phi,t]\bigr)
    \right],
\eeq
where $H_{\rm inf}\equiv H[\phi_{\rm max}]$, and $H[\phi]$ is the Hubble parameter including the axion potential.  
This equation applies when the $\ddot{\phi}$ term in the axion equation of motion is negligible.

The first term encodes the volume-weighting effect due to different expansion rates.  
$L$ is the comoving volume normalized with respect to $H_{\rm inf}$.  
If $L[\phi,t]$ stays constant, the physical volume grows as $e^{3H_{\rm inf} t}$.  
The terms inside the brackets describe the classical drift and the quantum diffusion contributions, respectively.

Neglecting the quantum diffusion term,\footnote{
For a quadratic potential with mass $m_\phi$, the timescale on which diffusion becomes relevant is  
$\sqrt{\Delta N_{\rm quantum}}\,H_{\rm inf}/(2\pi)=\sqrt{2/3}\,M_{\rm pl}$.  
The classical drift timescale is $\Delta N_{\rm classical}\sim 3H_{\rm inf}^2/m_\phi^2$.  
Requiring $\Delta N_{\rm quantum}<\Delta N_{\rm classical}$ gives  
$m_\phi > 3H_{\rm inf}^2/(2\sqrt{2}\pi M_{\rm pl})$.  
If $m_\phi$ is much smaller, diffusion balances expansion and produces a different eigen distribution.  
We do not consider that case because for $f_\phi \sim M_{\rm pl}$ one typically has  
$H_{\rm inf}\gg \L$.} 
we separate variables in $t$ and $\phi$ and consider the eigenvalue equation
\beq
 3\,(H[\phi]-H_{\rm inf})\,L[\phi,t]
  + \partial_\phi\!\left(
      \frac{V_{,\phi}}{3H[\phi]}\,L[\phi,t]\right)
      = E\,L[\phi,t].
\eeq
For $V_{,\phi} = -m_\phi^2(\phi-\phi_{\rm max})$, one finds that  
$E = -m_\phi^2/(3H_{\rm inf})$ is the largest eigenvalue compatible with a finite solution near $\phi\simeq \phi_{\rm max}$.  
This yields the dominant mode (see also \cite{Graham:2018jyp,Yin:2021uus})
\beq\laq{dis}
L[\phi,t]\;\propto\;
\exp\!\left[-\frac{m_\phi^2}{3H_{\rm inf}}\,t\right]\,
\exp\!\left[-\!\int^\phi\! d\phi'\,
\frac{3\,(H[\phi']-H_{\rm inf})}{V_{,\phi'}/(3H[\phi'])}\right]
\;\simeq\;
\exp\!\left[-\frac{m_\phi^2}{3H_{\rm inf}}\,t\right]\,
\exp\!\left[-\frac{3(\phi-\phi_{\rm max})^2}{4M_{\rm pl}^2}\right].
\eeq
In the last step we used
\beq
3\,(H[\phi']-H_{\rm inf})
\simeq \frac{V[\phi']-V[\phi_{\rm max}]}{2M_{\rm pl}^2 H_{\rm inf}}
\simeq -\frac{m_\phi^2(\phi'-\phi_{\rm max})^2}{4M_{\rm pl}^2 H_{\rm inf}}.
\eeq

Thus the distribution is peaked at the hilltop with variance  
$\sqrt{2/3}\,M_{\rm pl}$ for the quadratic potential.
When the effective decay constant $f_\phi$ (not $f_0$) is of order or larger than $M_{\rm pl}$, the single-branch axion potential is smooth over this variance, and the eigen distribution forms.  
If $f_\phi \ll M_{\rm pl}$, the variance exceeds the curvature scale of the potential and the field instead settles into the minimum.

Using $h[x]=1-\cos x$ and $f_\phi=\sqrt{2/3}\,M_{\rm pl}$, we have confirmed numerically by solving \Eq{FP} that the eigen distribution is stable.  
For smaller $f_\phi$, the distribution no longer stabilizes.

Now, let us include the effect of $\Gamma[\bar\theta]$.  
If for all values of $\bar\theta$ the condition  
\[
\Gamma[\bar\theta]/H_{\rm inf}^4 \times \Delta N_{\rm classical} \ll 1
\]
is satisfied, then the previous discussion remains unchanged.  
Here $\Delta N_{\rm classical}$ is the timescale for classical drift; for a quadratic hilltop potential with mass parameter $m_\phi^2$, one has  
\[
\Delta N_{\rm classical} \sim \frac{3H_{\rm inf}^2}{m_\phi^2}.
\]
Since the hilltop distribution forms on a timescale $\Delta N_{\rm classical}$, any decay that occurs more slowly than this does not distort the distribution.  
Although tunneling connects different branches, inflationary expansion drives each branch toward the same type of distribution as in \Eq{dis}.  
This corresponds to the regime we have assumed so far.

On the other hand, if for some range of $\bar\theta$ the opposite condition holds,
\[
\Gamma[\bar\theta]/H_{\rm inf}^4 \times \Delta N_{\rm classical} \gg 1,
\]
then the distribution \Eq{dis} is modified.  
Since $\Gamma[\bar\theta]$ is highly suppressed in the vicinity of the hilltop, the distribution becomes sharply localized around the hilltop with effective field-space boundaries; in particular, the variance becomes parametrically smaller than $\sqrt{2/3}\,M_{\rm pl}$.  
The mechanism is similar to that in the previous section: the hilltop, having a slightly larger expansion rate than the minimum, is statistically favored and dynamically selected.  

In this regime we again require $f_\phi \gtrsim M_{\rm pl}$.  
Because $\bar S \gtrsim 1$, the decay rate cannot be greatly enhanced, so tunneling cannot occurs extremely close to the hilltop.  
Focusing on the classical motion, the condition that the hilltop is selected before the field can roll out of the narrow region determined by the tunneling threshold is  
\beq 
\Delta N_{\rm classical} \times 
\frac{N^2 \Lambda^4}{6 M_{\rm pl}^2 H_{\rm inf}^2}
\;\sim\;
\frac{f_\phi^{\,2}}{M_{\rm pl}^{\,2}}
\;>\;1.
\eeq
We have also checked this behavior numerically by including the additional term  
$-\Gamma[\phi/f_\phi]/H_{\rm inf}^3 \, L[\phi,t]$  
on the right-hand side of \Eq{FP}, confirming that the distribution indeed becomes localized near the hilltop in this regime, by either using the quadratic hilltop or $h=1-\cos[x]$.

Therefore, when $f_\phi\gtrsim M_{\rm pl}$ the axion can be naturally placed near the hilltop.  We emphasize that this mechanism selects the hilltop through inflationary dynamics, 
independently of the detailed form of $h$, and does not require a super-Planckian 
(but not too small) fundamental decay constant $f_0$, provided that $N$ is sufficiently 
large.  Interestingly, a string axion with $f_0 \sim 10^{16\text{-}17}\,\GEV$ is a natural 
target regime in which this mechanism can operate.

After inflation, the averaged axion field value in our observable Universe lies  
close to $\bar\theta_{\rm max}$, with a $1\sigma$ spread determined by the variance obtained above,  
together with small fluctuations of order $\sqrt{\mathcal{O}(10)}\,H_{\rm inf}/(2\pi)$,  
where $\mathcal{O}(10)$ denotes the number of e-folds corresponding to our observable patch.

At late times, as a consequence of having $f_\phi \sim M_{\rm pl}$,  
the axion at the onset of oscillation can drive a mild secondary inflation or a brief supercooling phase,  
analogous to what occurs in a strongly first-order phase transition.  
A numerical solution of the homogeneous axion equation of motion with $h[x]=1-\cos x$ shows that,  
unless $\bar\theta_i \simeq \pi$ is finely tuned, the number of e-folds of this secondary inflation  
does not exceed $\mathcal{O}(1)$ for $f_\phi \lesssim 2M_{\rm pl}$, which is cosmologically safe.

In what follows, we consider the two representative cases in \Eq{cases} and focus on the  
post-inflationary epoch, independent of how the initial condition is prepared.  
For simplicity, we restrict our discussion to a two-state system, while keeping in mind that  
multiple branches may in principle trigger cascade tunneling.

\subsection{Post-inflationary evolution and onset of the transition}

After inflation ends, the axion field evolves according to
\beq
\ddot{\phi} + 3H\dot{\phi} = -V_{\phi}, \laq{KG}
\eeq
where $H=1/2t$ corresponds to the radiation-dominated epoch assuming instantaneous reheating for simplicity.  

By solving this equation, one can evaluate the time dependence of $\Gamma[\phi]$ and study the tunneling process. 
In particular, we introduce the standard parameters
\beq 
\alpha \equiv \frac{\epsilon}{\rho_{\rm rad}} 
   = \frac{\tl c\,\Lambda^4}{3H^2 M_{\rm pl}^2},
\qquad 
\beta \equiv \frac{d}{dt}\ln\Gamma,
\eeq 
where $\epsilon$ denotes the latent energy density.  
These parameters will be evaluated at $t=t_n$.  

If the tunneling completes well before the onset of axion oscillations, the discussion of the previous section applies.  
In this section, therefore, we focus on the case in which the relevant tunneling dynamics occurs when axion starts to evolve, 
\beq
m_\phi \sim H,
\eeq
during the radiation-dominated era, where $m_\phi \sim N\Lambda^2/f_\phi$ is the axion mass scale.  If we define the cosmic time for the equality, $t=t_{\rm osc}$, what we assume is $t_{\rm n}\sim t_{\rm osc}.$

This condition implies 
$\rho_{\rm rad} \sim m_\phi^2 f_\phi^2 \sim N^2 \Lambda^4$ 
for $f_\phi \sim M_{\rm pl}$, as motivated by string axions.  
Consequently, $\alpha \sim 1/N^2$, which is not particularly small unless $N$ is extremely large.

\subsubsection{$\bar{\theta}_i \lesssim  1$.}

\begin{figure*}[!t]
    \begin{center}
     \includegraphics[width=1\textwidth]{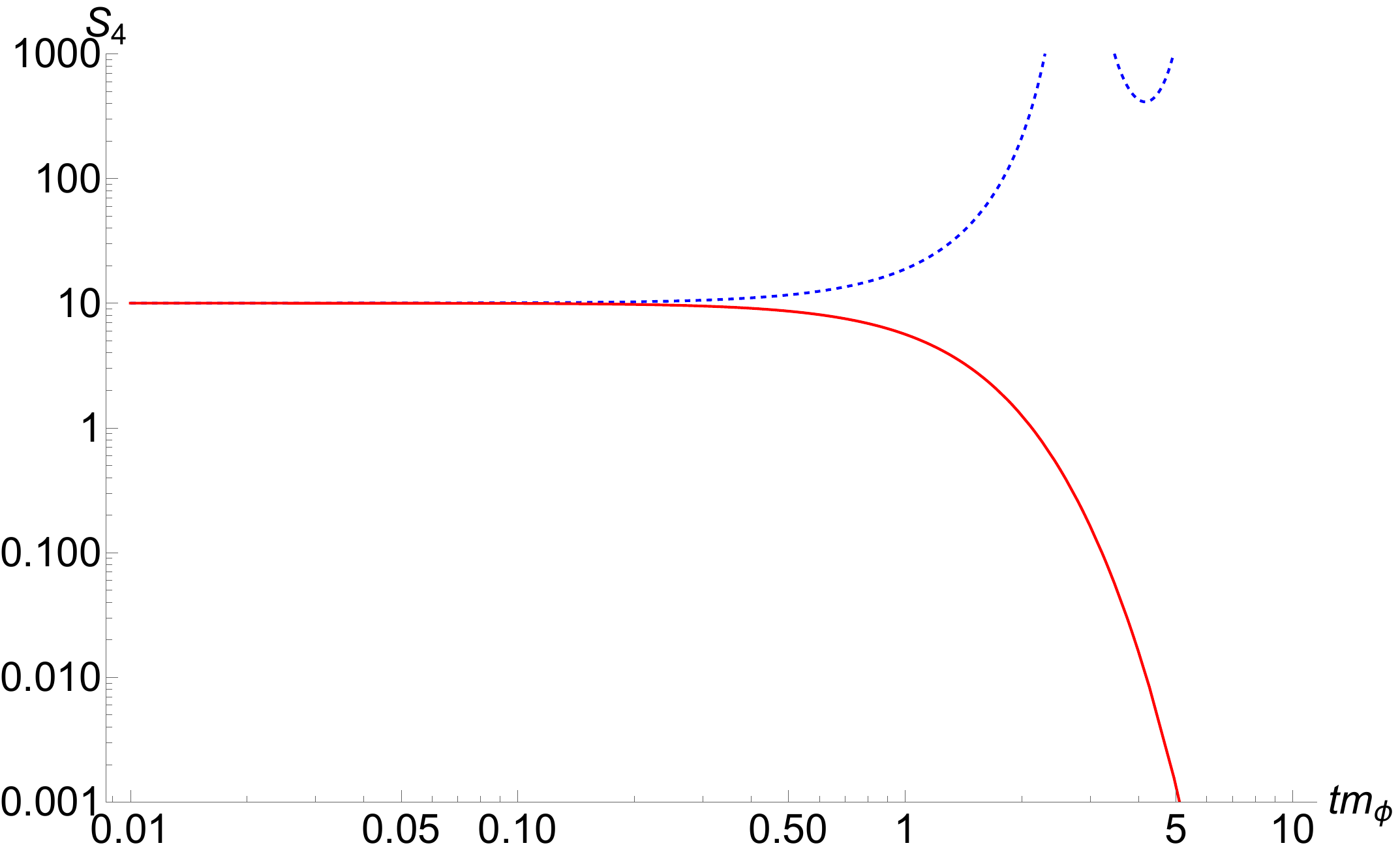}
        \vspace{-15mm}
    \end{center}
    \caption{
    Time evolution of $S_4$, which governs the bubble nucleation rate in Eq.~(\ref{rate1}). 
    The blue dashed (red solid) curve shows the result with $|\bar\theta_i|\lesssim 1$ ($\bar{\theta}_i\sim \bar\theta_\text{max}$) with $\bar\theta_i^{-3} \bar{S}=10$ $(\delta\bar\theta_i^{-3} \bar{S}=10)$. A radiation-dominated Universe is assumed.}
    \label{fig:1}
\end{figure*}

When $\theta_i/N \lesssim 1$, the potential is approximately
\beq
V \simeq \frac{1}{2}m_\phi^2 \phi^2, \laq{vapp0}
\eeq
and the tunneling rate between neighboring branches is given by\footnote{Again, the formula here cannot be applied for $-\pi < \theta_i < \pi$, since the decay rate in this region vanishes exactly. 
For illustrative purposes, we assume that $\theta_i$ is sufficiently far from $\pi$ while still satisfying $\bar\theta_i \lesssim 1$.}
\beq \label{rate1}
\,S_4=\frac{\bar{S}}{|\phi(t)/f_\phi|^3}.
\eeq
{In the radiation-dominated epoch, i.e., $H=1/2t$, \Eq{KG} can be analytically solved as
\beq 
    \phi(t) = \phi_i\left(\frac{2}{m_\phi t}\right)^{1/4}\Gamma\left(\frac{5}{4}\right)J_{1/4}(m_\phi t),
\eeq
where $\phi_i$ is the initial value of $\phi$, $\Gamma(x)$ is the gamma function, and $J_{\alpha}(x)$ is the Bessel function of the first kind.
} The resulting $S_4$ by taking $\bar{S}\theta_i^{-3}=10$ is shown by the blue dashed line in Fig.~\ref{fig:1} by varying $m_\f t$. 

While $H=1/2t \gg m_\phi$, the axion is overdamped and effectively frozen.  
If tunneling occurs during this stage, the discussion reduces to that of the previous section.

When $H \lesssim m_\phi$, the field begins to oscillate around the potential minimum.  
Since the oscillation frequency $m_\phi \sim N\Lambda^2/f_\phi$ is much smaller than the characteristic bounce scale $\sim \Lambda$ for sufficiently large $f_\phi$, 
we can evaluate the tunneling rate adiabatically as a function of $\theta(t)$.  
After one oscillation, the maximal exponent, corresponding to the oscillation peak in Fig.\ref{fig:1} is found to be
$
S_4\approx 411 \left(\frac{\bar{S}\,\bar{\theta}_i^{-3}}{10}\right),
$
which is already too large for tunneling to occur within the age of the (radiation-dominated) Universe.

Given that $\bar{S}/|\bar{\theta}_i|^3 \gg 1$ for $|\bar{\theta}_i|\lesssim 1$ and $\bar{S}\gg1$, 
it is unlikely to realize $\bar{S}\bar{\theta}_i^{-3} < 10$.  
Thus, the tunneling can only occur before the first oscillation, i.e.\ during the slow-roll phase.

During the slow roll, the solution in a radiation-dominated Universe is well approximated by
\beq \label{ana}
\phi(t) = \bar{\theta}_i f_\phi e^{-t^2 m_\phi^2/3},
\eeq
and the nucleation rate decreases rapidly as $|\phi/f_\phi|$ diminishes.  
In this regime we find
\beq
\beta \approx 
-200\,H_n\,\frac{e^{m_\phi^2 t_n^2}\,\bar{S}\,\theta_i^{-3}}{100}\,
\frac{t_n m_\phi^2}{H_n}.
\eeq

Given that $t_n m_\phi^2/H_n \sim 1$, corresponding to the onset of oscillations, this yields a large negative $\beta$.  
Hence bubble nucleation can start but soon shuts off.  
There is no problem of false-vacuum inflation here, because the axion in the ``false vacuum" starts to oscillate around the minimum and this can only be the slow-roll inflation, which does not last long if $f_\f\lesssim M_{\rm pl}.$
The resulting bubble wall between the ``false vacuum" and ``true vacuum" after the oscillation is more akin to a closed domain-wall loop, which later collapses due to its tension.

We also comment that the condition $t_{\rm n} \sim t_{\rm osc}$ in the case $\bar\theta_i \lesssim 1$ requires tuning, 
since it is not dynamically determined when $\beta < 0$.  
In contrast, as we will discuss, for the case $\bar\theta_i \sim \bar \theta_{\rm max}$ this condition is instead driven dynamically.

\paragraph{The fate of the bubble wall with $\bar\theta_i\lesssim 1$}
There are two possible scenarios for the fate of the bubble walls.  
If the condition $\Gamma \sim H^4$ is satisfied when the phase transition occurs and the axion oscillation starts sufficiently late, 
the walls are strongly accelerated.  
In this case, although the nucleation rate eventually decreases (corresponding to a negative $\beta$), 
each bubble expands until its radius becomes of order the Hubble scale.  
Since there are $\mathcal{O}(1)$ bubbles per Hubble volume, they still collide and the tunneling effectively completes. We may call this is a first order phase transition. 
Due to the absence of the friction force, the wall is ultra-relativistic. 

The second possibility is that the acceleration is not efficient enough, or that $\Gamma \lesssim H^4$ at the onset of oscillation, so that bubble formation is a rare event.  
In this case, a rarely formed bubble wall, after being accelerated during the slow-roll phase, expands until it loses its kinetic energy due to the redshift from the cosmic expansion.  
Neglecting the redshift, 
the kinetic energy of the wall is estimated as
\beq
E_{\rm wall,kin} \sim \frac{4\pi}{3} R_{\rm osc}^3\,\epsilon,
\eeq
where $R_{\rm osc}$ is the bubble radius when the potential difference disappears due to the oscillation of the axion in the ``false'' vacuum.  
This is much larger than the tension energy of a static bubble with the same radius,
\beq
E_{\rm wall,rest} \sim 4\pi \sigma R_{\rm osc}^2,
\eeq
since $\sigma \sim \Lambda^3$, $\epsilon \sim \Lambda^4$, $R_{\rm osc} \sim 1/H_{\rm osc}={2t_{\rm osc}}$, and the Hubble radius $1/H_{\rm osc}$ is a cosmological scale much larger than $1/\Lambda$.

Because the wall is ultra-relativistic, its kinetic energy redshifts as $E_{\rm wall,kin}\propto a^{-1}$ (see \cite{Boehm:2002bm} for the radiation equation of state of the relativistic wall), 
whereas the would-be rest energy of a horizon-sized wall scales as $E_{\rm wall,rest}\propto a^{4}$ in a radiation-dominated Universe 
(due to $R\sim H^{-1}\propto a^{2}$).  
The scale factor at which the wall becomes non-relativistic and starts to collapse is then estimated  by equating them as
\beq 
\frac{a_{\rm nr}}{a_{\rm osc}}
\;\sim\;
\left(\frac{E_{\rm wall,kin}}{E_{\rm wall,rest}}\right)^{1/5}
\sim  
\left(\frac{\Lambda}{3H_{\rm osc}}\right)^{1/5}.
\eeq 
Since the wall velocity cannot exceed the speed of light, its radius remains of order the Hubble horizon until $a = a_{\rm nr}$.  
After that, the wall begins to shrink due to its tension and collapses within an $\mathcal{O}(1)$ Hubble time.  
From energy conservation, the energy carried by the wall just before collapse is
\beq 
E_{\rm wall,collapse}
\sim 
E_{\rm wall,kin}\left(\frac{a_{\rm nr}}{a_{\rm osc}}\right)^{-1}
\sim 
\frac{4\pi}{3}\,
\frac{\Lambda^{21/5}}{H_{\rm osc}^{15/5}}.
\eeq 
This energy is eventually converted into Yang-Mills particles and axions. 
Since the wall configuration is (approximately) spherically symmetric, the emission of gravitational waves, which requires a time-dependent quadrupole moment, is expected to be strongly suppressed. 

Note that the total radiation energy inside a Hubble volume is 
$\rho_{\rm rad} H^{-3}\propto a^{2}$, 
which is typically larger than the wall energy, even if $\rho_{\rm rad}\sim N^2\Lambda^4$ at the time of tunneling.  
However, if $\Lambda$ is sufficiently high, the wall energy can become comparable to the radiation energy over a finite epoch, potentially leading to large overdensities.

The resulting overdensities may also seed axion miniclusters or primordial black holes (see, e.g., Ref.~\cite{Hogan:1988mp,Kolb:1993zz,Kolb:1993hw,Kolb:1994fi,Kolb:1995bu, Jedamzik:1999am,Murai:2025hse,Kitajima:2025shn}).

\paragraph{Axion dynamics with bubble wall}

So far we focused on the homogeneous axion oscillation let us study the axion wave dynamics associated with the bubble wall. 

After the onset of oscillation, an axion domain wall can be induced by the bubble wall~\cite{Lee:2024xjb}, because the minima of $\theta$ differ between the branches, and axion wave can be generated because the motion of the (induced) wall \cite{Lee:2024oaz}.

To simulate the system, we solve the equation of motion for spherically symmetric $\f$,
\beq
\left(\partial_t^2  -\frac{1}{r^2}\partial_r\!\left(r^2 \partial_r\right) + m_\phi^2\right)\phi
 - m_\phi^2\, v\, \Theta\!\left[-(r - v_w t)\right] = 0.
\eeq
Here, $v$ denotes the difference in the vacuum expectation values between the two adjacent vacua.  
In the large-$N$ limit, it is approximately given by $v \approx 2\pi\,f_\phi/N$.
In our scenario, a spherically symmetric bubble and a homogeneous axion background are obtained, the approximation of rotational symmetry in this equation is well justified. 

\begin{figure*}[!t]
    \begin{center}
     \includegraphics[width=0.49\textwidth]{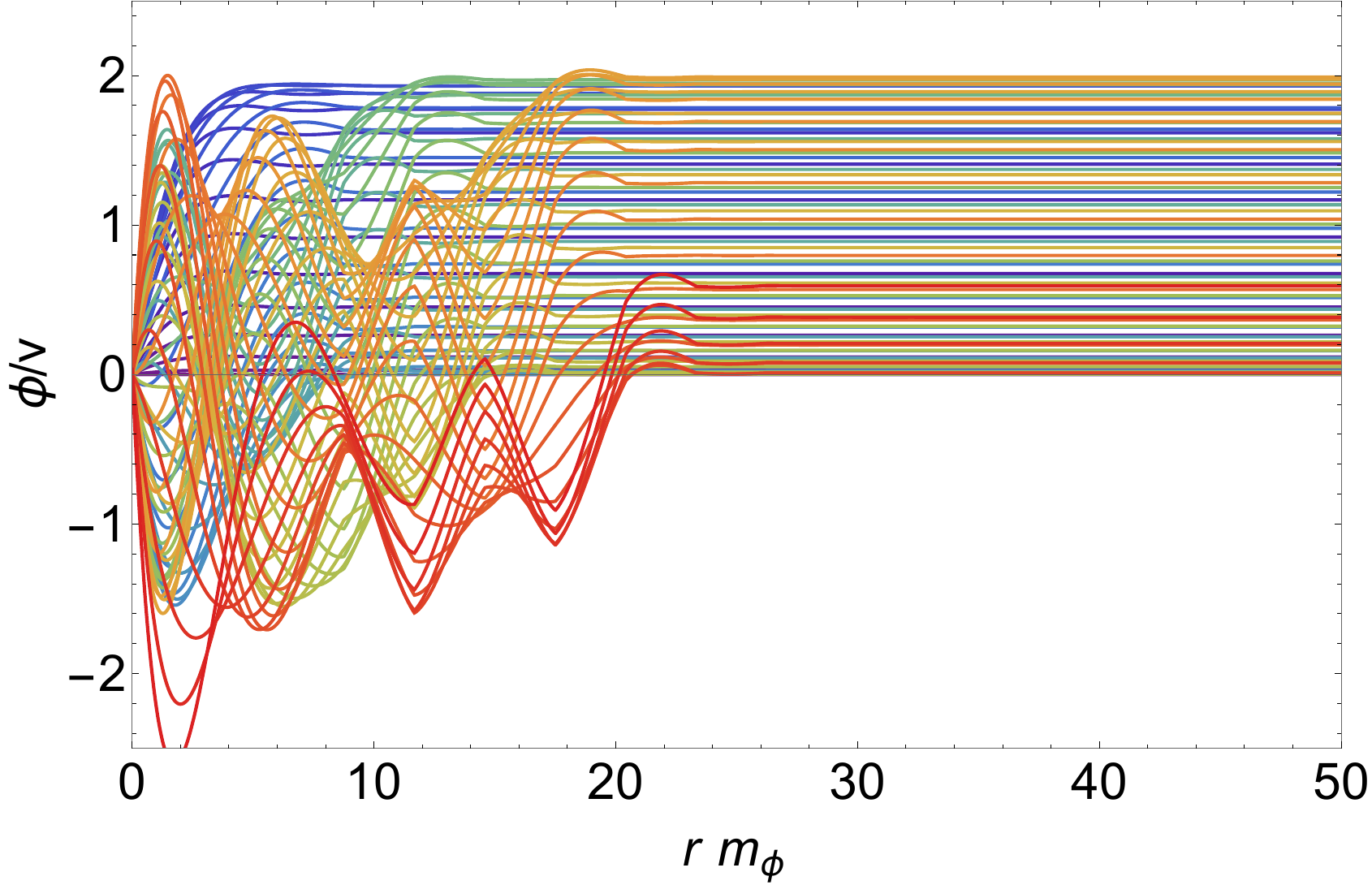}
     \includegraphics[width=0.49\textwidth]{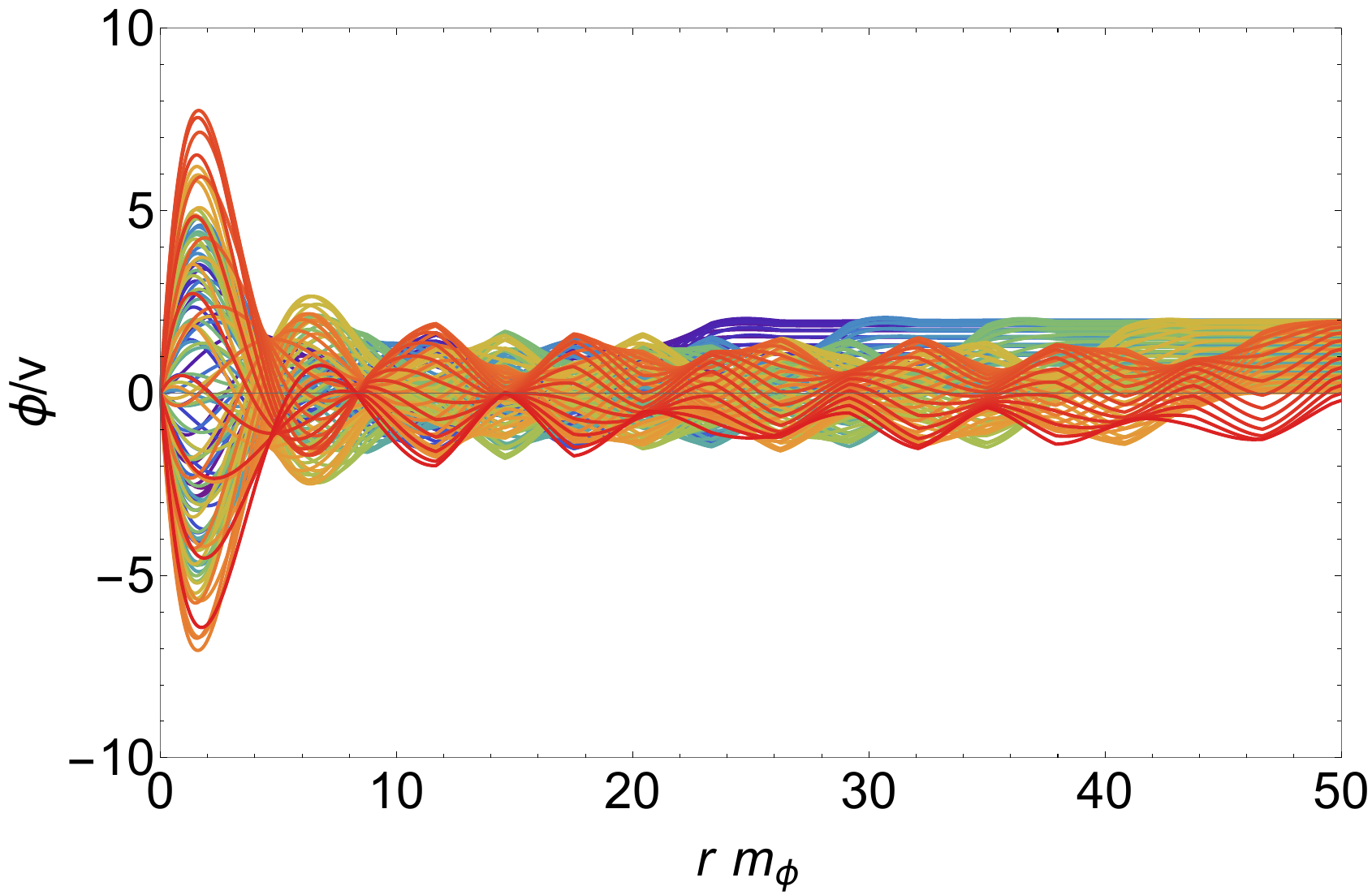}
    \end{center}
    \caption{
    The simulation of the axion field configuration at different $t$ with the wall background. The left panel denotes the result with smaller $t=(0-20) /m_\f$ while the right panel denotes the configuration at the $t=(20-50) /m_\f$. The wall sweeps the whole area from $t=0-50/m_\f$.}
    \label{fig:axionwave}
\end{figure*}

The results are shown in Figs.~\ref{fig:axionwave}. 
We take $v_w\approx 1$, and the precise value does not change the qualitative behavior. 
One can see that outside the wall, the axion undergoes the usual coherent oscillation. 
Since the minima of the oscillation in the two regions are different, as shown in the left panel of the figure, a domain wall in the axion field is formed. 
When the axion condensate flows into the bubble due to the expansion, (approximately) stationary axion waves with characteristic wave number $\sim 1/m_\phi$ are generated. In particular, these waves are enhanced when they converge near the center of the bubble. 

We have also checked that the axion wave remains even when the bubble bounces and collapses. 
This is because the axion mass is the same in both branches, so the axion wave can freely cross the wall while remaining connected to the induced domain wall (see \cite{Lee:2024xjb,Lee:2024oaz}).
For the same reason, the shock wave outside the bubble, which was found in \cite{Lee:2024oaz} for a usual phase transition with different vacuum energies, is suppressed in our scenario. 

Thus, in the scenario with $\bar\theta_i\lesssim 1$ where the axion constitutes the dark matter, this wave could act as a seed for axion clumps.

\subsubsection{$
\bar\theta_i \sim \bar \theta_{\rm max}$}

Let $\delta\bar{\theta}_i \equiv \bar\theta_{\rm max} - \bar{\theta}_i$ with $\delta\bar{\theta}_i > 0$ without loss of generality.  
Near the hilltop, the potential can be approximated by
\beq
V \simeq -\frac{1}{2}m_\phi^2(\phi - \bar\theta_{\rm max} f_\phi)^2 \;\equiv\; -\frac{1}{2}m_\phi^2 (\delta\phi)^2,
\eeq
and the tunneling rate between neighboring branches is given by
\beq
\Gamma(t) \propto \exp\!\left[-\,\frac{\bar{S}}{|\delta\phi/f_\phi|^3}\right].
\eeq
We note that $m_\f$ can differ from the vacuum mass by an $\O(1)$ factor, since the curvature at the hilltop does not necessarily coincide with the curvature of the potential in the vacuum. Taking this possible difference into account does not modify our conclusions, because the late-time oscillation is not important for the phase transition phenomena discussed in this section.

Also in this case, \Eq{KG} can be analytically solved and the solution is 
\beq 
    \delta\phi(t) = \delta\bar\theta_if_\phi\left(\frac{2}{m_\phi t}\right)^{1/4}\Gamma\left(\frac{5}{4}\right)I_{1/4}(m_\phi t), \laq{detaphit}
\eeq
where $I_{\alpha}(x)$ is the modified Bessel function of the first kind. The resulting $S_4$  by taking $\bar{S}\delta\theta_i^{-3}=10$ is shown by the red solid line in  Fig.~\ref{fig:1}.

Interestingly, in this regime the tunneling rate grows with time.  
This behavior resembles that of a conventional first-order phase transition, provided that bubbles nucleate and collide during the period when the exponent is increasing.  
Indeed,
\beq 
\beta \;\approx\;
-\!\left(\dt\frac{\bar{S}}{(\delta\phi/f_\phi)^3}\right)\Big|_{t=t_n}
= -\left.\frac{\bar{S}}{(\delta\phi/f_\phi)^3}\,\frac{d}{dt}\ln(\delta\phi^{-3})\right|_{t=t_n}
\;\approx\;
{300m_\phi\times \frac{\bar{S}\,(\delta\phi/f_\phi)^{-3}|_{t=t_n}}{100}},
\eeq
for sufficiently large $t m_\phi$, where we used that $-\dt\ln(\delta\phi^{-3})$ approaches {$3m_\phi$} (see \Eq{detaphit}).

Before the onset of axion oscillations, the potential difference between neighboring branches is $\epsilon \sim N\,\Lambda^4 h' 2\pi$.  
Since the sector is sequestered, there are no background particles to exert friction on the bubble walls.  
Consequently, the walls are accelerated to ultra-relativistic velocities, $v_w \simeq 1$, until they collide with each other.  
The typical bubble size at collision is approximately
\beq
R_{\rm coll} \sim v_w\,\beta^{-1}
   \sim \frac{1}{300}\,
   \frac{100}{\bar{S}(\delta\phi/f_\phi)^{-3}|_{t=t_n}}\,
   H_n^{-1}.
\eeq
The violent collisions of such ultra-relativistic domain walls release latent energy of order $\epsilon$ and can efficiently source stochastic gravitational waves.  
The resulting spectrum can be estimated using the standard envelope approximation or the more recent numerical analyses in e.g. Ref.~\cite{Caprini:2019egz}, 
taking into account that the present scenario corresponds to the vacuum-dominated, frictionless limit of a strongly first-order transition.

On the other hand, with the multi-tunneling formation mentioned close to Fig.\ref{fig:binb}, the estimation may differ and warrant future study.

Finally, we note again that the parameter $\alpha$ can become significantly large when the axion is string-motivated with a fundamental decay constant, $f_0$, close to the Planck scale.  
Since the nucleation time is determined by the condition $H \sim m_\phi$, the vacuum energy difference $\Lambda^4$ can be comparable to the radiation energy density, leading to $\alpha$ as large as  $\mathcal{O}(1/N^2)$.

\section{Conclusions and discussion}

In this work, we have investigated the cosmological implications of the multi-branch vacuum structure of a Yang-Mills theory coupled to an axion, 
under the assumption that the Yang-Mills sector is fully sequestered and is not reheated.  
Even in such a minimal setup, the axion inherits a potential that is directly tied to the 
tunneling rate between adjacent Yang-Mills branches, giving rise to qualitatively new phenomena 
that are absent in conventional axion cosmology.

First, we showed that inflationary dynamics naturally select particular Yang-Mills branches.    
In this regime, large regions of the post-inflationary Universe remain trapped in highly excited branches, and the axion field value after inflation can be highly nontrivial.

Second, we demonstrated that the axion dynamics after inflation can affect and even trigger transitions between 
Yang-Mills branches, generating a new class of strongly first-order phase transitions.  
These axion-induced transitions differ sharply from conventional thermal phase transitions:  
the bubble walls experience essentially no friction, become ultra-relativistic,  
and can induce rich dynamics including bouncing bubbles and nested 
``bubbles-within-bubbles.''  
Such cascade behavior is particularly pronounced near the hilltop, 
where the tunneling rate grows rapidly as the axion rolls.

Third, we identified several potential observational consequences.  
If the relevant Yang-Mills scale is not too small compared with the radiation energy density at 
the transition, the ultra-relativistic bubble collisions can yield sizable stochastic 
gravitational-wave backgrounds.  
In cases where bubble formation is rare and the potential bias disappears as the axion begins to oscillate, 
bouncing bubbles can generate large overdensities, potentially leading to the formation of primordial 
black holes or significant particle production within the Yang-Mills sector.  
For axions with decay constants near the Planck scale, as motivated by the Axiverse,  
these effects become especially pronounced.

Overall, our results show that the multi-branch structure inherent to Yang-Mills theories can have 
a dramatic impact on axion cosmology, even in the absence of thermal population or additional 
hidden-sector dynamics.  
These effects represent novel predictions of the Axiverse and highlight the importance of accounting for 
Yang-Mills multi-branch structure in realistic axion models.  
Conversely, obtaining the standard picture of axion cosmology requires suppressing or eliminating these 
phenomena-for example, by breaking the Yang-Mills gauge group at high energies inducing the potential via small instanton or by considering 
theories without a multi-branch vacuum structure.
\\

The phenomena found here should have various interesting applications. For instance, a hilltop axion that starts oscillating around the present Universe is mildly favored in light of the recent DESI result and isotropic cosmic birefringence~\cite{Takahashi:2020tqv,Tada:2024znt,Kitajima:2022jzz,Gonzalez:2022mcx,Lee:2025yvn,Nakagawa:2025ejs}. In such a case, the phase transition associated with the Yang-Mills multi-branch structure may arise as an accompanied prediction.

The bubble formation is around the present Universe, and walls can interact with minerals in the Earth in the early Universe through weakly coupled higher-dimensional operators, potentially leaving observable damages in the ``Paleo detectors"~\cite{Yin:2025wuv} (see also~\cite{Baum:2023cct,Hirose:2025jht}). 

Given the higher dimensional term, even if it is suppressed, the ultra-relativistic bubbles in the hot early Universe can interact with the Standard Model particles to produce energetic beyond standard model particles which can be relevant to the cold or warm dark matter production~\cite{Azatov:2021ifm,Baldes:2022oev,Azatov:2022tii,Azatov:2023xem,Azatov:2024crd,Ai:2024ikj} or baryogenesis~\cite{Azatov:2021irb,Baldes:2021vyz,Azatov:2023xem,Chun:2023ezg}. Although most of the references assume that the bubble Higgs is thermalized, the assumption is not very relevant for the estimation of the particle production.

As mentioned above, a precise understanding of the gravitational-wave spectrum is important for probing fundamental theories that exhibit a multi-branch feature in the Yang-Mills sector. 

It is also worthwhile to study the axion abundance and structure formation while taking into account the aforementioned phenomena in the context of axion dark matter, since our setup may serve as one of the minimal realizations of axion dark matter models.

\section*{Acknowledgement}
W.Y. is supported by JSPS KAKENHI Grant Nos.  22K14029, 22H01215, and Selective Research Fund from Tokyo Metropolitan University.
T.S. is supported by Graduate Program on Physics for the Universe (GP-PU)  and AGS RISE Program from Tohoku University.

\bibliography{bib.bib}
\end{document}